\documentclass{jfm}
\usepackage{graphicx}
\usepackage{newtxtext}
\usepackage{newtxmath}
\usepackage{natbib}
\usepackage{siunitx}
\usepackage{amsmath}
\usepackage{xcolor}
\usepackage{hyperref}
\hypersetup{
    colorlinks = true,
    urlcolor   = blue,
    citecolor  = black,
}

\newcommand{\RomanNumeralCaps}[1]

\title{{Hydrodynamic roughness induced by a multiscale topography}}

\author{
Pan Jia\aff{1}
  \corresp{\email{jiapan@hit.edu.cn}},
Bruno Andreotti\aff{2}
\and
Philippe Claudin\aff{3}
}

\affiliation{
\aff{1}School of Science, Harbin Institute of Technology, 518055 Shenzhen, P. R. China 
\aff{2}Laboratoire de Physique de l'ENS, UMR 8550 Ecole Normale Sup{\'e}rieure -- CNRS -- Universit{\'e} PSL -- Universit{\'e} Paris Cit\'e -- Sorbonne Universit{\'e}, 24 rue Lhomond, 75005 Paris, France
\aff{3}Physique et M{\'e}canique des Milieux H{\'e}t{\'e}rog{\`e}nes, UMR 7636 CNRS -- ESPCI Paris -- Universit{\'e} PSL -- Universit{\'e} Paris Cit\'e -- Sorbonne Universit{\'e}, 10 rue Vauquelin, 75005 Paris, France
}

\begin{document}
\maketitle

\begin{abstract}
Turbulent flows above a solid surface are characterised by a hydrodynamic roughness that represents, for the far velocity field, the typical length scale at which momentum mixing occurs close to the surface. Here, we are theoretically interested in the hydrodynamic roughness induced by a two-dimensional modulated surface, {the elevation profile of which} is decomposed in Fourier {modes}. We describe the flow for a sinusoidal mode of given wavelength and amplitude with RANS equations closed by means of a mixing-length approach that takes into account a possible surface geometrical roughness as well as the presence of a viscous sublayer. It also incorporates spatial transient effects at the laminar-turbulent transition. Performing a weekly non-linear expansion in the bedform aspect ratio, we predict the effective hydrodynamic roughness when the surface wavelength is varied and we show that it presents a non-monotonic behaviour at the laminar-turbulent transition when the surface is {hydrodynamically} smooth.  {Further,} with a self-consistent looped calculation, we are able to recover the smooth-rough transition of a flat surface, for which the hydrodynamic roughness changes from a regime where it is dominated by the viscous length to another one where it scales with the surface corrugation. We finally apply the results to natural patterns resulted from hydrodynamic instabilities such as those associated with dissolution or sediment transport. We discuss in particular the aspect ratio selection of dissolution bedforms and roughness hierarchy in superimposed ripples and dunes.
\end{abstract}


\section{Introduction}
\label{sec:intro}

The length scale over which mixing occurs in the superficial layer of a turbulent boundary layer, close to a solid surface, is a key quantity in meteorology and geomorphology. This length, later defined as the `hydrodynamic roughness', is for example necessary to account for the effect of relief and vegetation on winds \citep{gillies2007shear, king2006aeolian, lancaster1998influence, marticorena1995modeling, raupach1992drag, raupach1993effect, wolfe1993protective, finnigan1988air, bradley1980experimental}. Remote sensing methods including satellite imagery \citep{jasinski1999estimation}, airborne LIDAR \citep{paul2013estimation}, tethersonde and eddy correlation systems \citep{tsai2010measurements, tsuang2003determining}, GPS radiosondes \citep{han2015estimates}, ERS scattermeter \citep{prigent2005estimation}, and terrestrial laser scanning \citep{nield2013estimating}, have been employed to evaluate the hydrodynamic roughness and its relation to geometrical asperities of the ground. In glaciology, these measurements are also used to quantify the turbulent heat exchange between the surface of glaciers and the atmosphere, and their contribution to the ice melt \citep{smeets2008temporal}. Similarly, the determination of the hydrodynamic roughness of the sea surface is important in oceanography to quantify the exchange of momentum at the air-sea interface \citep{maat1991roughness}  {and the nucleation of waves} \citep{perrard2019turbulent, aulnette2019wind}. As a final example, this length scale directly controls the vertical thermal exchange with the atmosphere in cities \citep{kent2017aerodynamic, crago2012equations}.

 {Some} of these applied problems belong to the generic case of a turbulent  {boundary layer above a flat rough surface}. Since the earliest investigations in the 1930s, including those of \citet{nikuradse1950lawsA} and \citet{schlichting1937experimentalA}, it has been found that the velocity profile in such flows is logarithmic with respect to the distance $z$ to the surface. Following dimensional analysis \citep{prandtl1925,schlichting2000boundary,pope2000turbulent}, the rate of momentum mixing in this turbulent layer is determined by the mean velocity gradient, over a length scale, called the mixing length, $\ell$, proportional to the geometrical distance $z$. For a steady and homogenous flow over a flat substrate in neutral conditions, the velocity profile away from the boundary takes the classical form:
\begin{equation}
u_x = \frac{u_*}{\kappa}\ln\left(\frac{z}{z_0}\right)\,,
\label{LogProfile}
\end{equation}
where $u_x$ is the time-averaged velocity along the flow direction $x$. $u_*$ is the friction velocity, related by $u_* =  \sqrt{\tau_{xz}/\rho}$ to the shear stress $\tau_{xz}$ and the fluid density $\rho$. $\tau_{xz}$ represents the flux of momentum across surfaces parallel to the solid substrate and is therefore conserved in the $z$ direction. $\kappa \approx 0.4$ is the phenomenological von K{\'a}rm{\'a}n constant. { The length $z_0$ is the altitude at which the logarithmic velocity profile (\ref{LogProfile}) seems to vanish above the ground, when extrapolated, hence appearing as a \emph{reference level} below which the velocity becomes very small.} It is called the {hydraulic roughness in hydraulic engineering} \citep{van1984sediment,flack2010review}, and the aerodynamic roughness in the case of an air flow, and here generically named hydrodynamic roughness. $z_0$ can be interpreted as a mixing length governing the turbulent fluctuations close to the solid surface.  {The physical origin of the hydrodynamic roughness depends on the processes at play close to the surface. It has for instance been experimentally studied in wind tunnels with discrete roughness elements placed on a flat substrate \citep{taylor1976some,schmid1995influence,hobson1999large,sadique2017aerodynamic}. It is modified when the solid is covered by slender flexible structures such as plants \citep{de2008effects}. It can be produced by turbulence fluctuations themselves, by roughening of the surface of the ocean. Another example is the case of sediment transport, which can dominate the interaction between the granular bed surface and the flow, and typically increases the hydrodynamic roughness \citep{owen1964saltation,raupach1991saltation,gillette1998change,sherman2008aerodynamic, duran2011aeolian}.}

\begin{figure}
\centering
\includegraphics[scale=1.0]{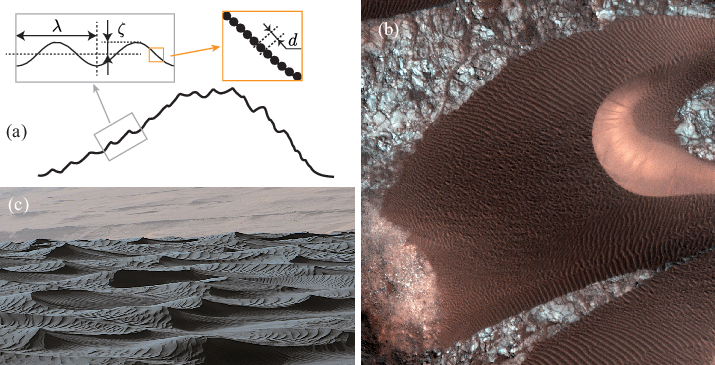}
\caption{ (a) Schematic of superimposed bedforms: grains of size $d$ form sand ripples of wavelength $\lambda$ and amplitude $\zeta$, which themselves cover the surface of a dune; the hydrodynamic roughness at the scale of the large bedform is induced by the small ones. (b) Martian dune of Nili Patera dune field (from HIRISE picture ESP\underline{\ }062069\underline{\ }1890) with superimposed meter-scale ripples. Dune size $\simeq 300$~m. (c) Photograph of the surface of a Martian dune in Gale Crater by NASA's Curiosity rover (Image Credit: NASA/JPL-Caltech/MSSS PIA20755). Small decimetric ripples superimposed on large meter-scale bedforms.}
\label{multiscale}
\end{figure}

 {A situation of interest is that of a turbulent flow over an elevation profile that presents several nested length scales. This situation of a gentle topography with superimposed patterns is typically relevant of sedimentary bedforms  \citep{van1982equivalent, van1984sediment,venditti2005morphodynamics,elbelrhiti2005field,narteau2009setting,nield2013estimating,duran2019unified}. A generic example is grains that form sand ripples, themselves covering the surface of dunes. As shown in figure~\ref{multiscale}, sedimentary bedforms can present up to four interlocking scales: grains, small impact ripples, large hydrodynamic ripples and dunes. How to account for the fact that the grains control the hydrodynamic roughness for the flow over the sand ripples, and the ripples that over the dunes?  How to handle such a multi-scale roughness hierarchy? Here, we consider a two-scale problem, and investigate the hydrodynamic roughness induced by a large scale topography whose surface presents a roughness at a much smaller scale. What is the effect of the \emph{inner} surface roughness on the \emph{outer} topography-induced roughness, far from the surface? What are the separate effects of amplitude and horizontal length scale of the large scale elevation profile on the outer hydrodynamic roughness? Is there a signature of the laminar-turbulent transition on this hydrodynamic roughness?}

 {The article is organised as follows. In section~\ref{flowdescription}, we perform the weakly nonlinear analysis of the turbulent flow above a modulated surface using two expansion techniques. We first discuss the smooth-rough transition, and present the turbulent closure used to relate the Reynolds stresses to the velocity gradient inspired from models developed for erosion, dissolution and sublimation pattern formation \citep{charru2013sand, claudin2017dissolution, duran2019unified}. This closure involves a mixing length which allows us not only to account for surface geometrical roughness and viscous sublayer, but also to incorporate transient effects at the laminar-turbulent transition. The most technical mathematical considerations are gathered in several appendices. The results are presented in section~\ref{TopographyInducedHydrodynamicRoughness} where we study the outer hydrodynamic roughness induced by a topography, with an emphasis on the effect of the inner, surface roughness. We show that a self-consistent, looped calculation, allows us to recover the hydrodynamic roughness induced by a rough surface of equivalent grain size $d$.  We then come back to the questions mentioned in the previous paragraph, compare some of our findings with salient results  {from} the literature, and address the laminar-turbulent transition from the point of view of bed roughness. Open problems and perspectives conclude the manuscript.}

\begin{figure}
\centering
\includegraphics[scale=1.0]{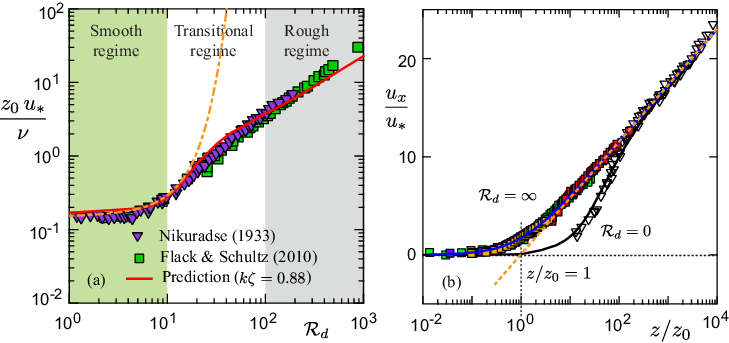}
\caption{(a) Hydrodynamic roughness $z_0$ measurements, including \citet{flack2010review} (green squares) and those by Nikuradse (1933) (violet triangles), as a function of the dimensionless equivalent grain size $\mathcal{R}_d = d u_*/\nu$. Green background $\mathcal{R}_d \lesssim 10$: smooth regime for which $z_0 \simeq \nu/(7u_*)$. Gray background $\mathcal{R}_d \gtrsim 10^2$: rough regime for which $z_0 \simeq d/30$. Lines: theoretical prediction of an effective roughness $z_e$ induced by a sinusoidal surface of wavelength $\lambda=d$ and amplitude $\zeta = 0.88 d/2\pi$ (see section~\ref{Matching}). Dash-dotted orange line: perturbative calculation (Eq. \ref{zedirect}). Solid red line: self-consistent calculation. (b) Dimensionless velocity profiles $u_x/u_*$ from data over a smooth flat surface \citet{johnson1989active, wei1989reynolds} (inverted triangles) and over various rough flat surfaces \citep{schultz2009turbulent,flack2010review} (squares). Lines: function $\mathcal{U}$ (Eq.~\ref{EqDiffmathcalU}) in the smooth (black curve, $\mathcal{R}_d = 0 $) and rough (blue curve, $\mathcal{R}_d \to \infty $) limits. The distance to the surface $z$ is rescaled by $z_0$, deduced by an extrapolation to a vanishing value of the profile semi-log fit far from the surface (dashed orange line).}
\label{schematic}
\end{figure}

\section{Hydrodynamic model}
\label{flowdescription}

 {\subsection{Smooth rough transition}
We consider here a turbulent boundary layer above a solid surface whose elevation profile presents two well-separated length scales:  a large-scale topography whose surface presents a roughness at a smaller scale. This situation is typically that of sand grains at the surface of sand ripples, or that of sand ripples at the surface of sand dunes. A simplifying description is to consider the large scale topography as smooth, described by a vertical elevation profile $z=Z(x)$, invariant along the spanwise $y$-direction, and to take into account the effect of the roughness of the solid surface in the hydrodynamic model (figure~\ref{multiscale}a).}

 {The situation where the topography is flat provides important constrains to calibrate such a model (figure~\ref{schematic}). When the surface roughness is much smaller than the thickness of the viscous sublayer $\sim \nu/u_*$, where $\nu$ is the kinematic viscosity of the fluid, the hydrodynamic regime is said \emph{smooth}: the logarithmic velocity profile is regularised by a laminar flow close to the surface \citep{schlichting2000boundary, pope2000turbulent}. In this case, the hydrodynamic roughness is consequently found to be controlled by the viscous length $\nu/u_*$, independent of the amplitude of the solid surface asperities and is around $z_0 \sim  \nu/(7u_*)$ \citep{Raupach1991,schlichting2000boundary}. In contrast, when the viscous sublayer is thin enough, $z_0$ is rather controlled by the geometrical roughness of the solid surface and the hydrodynamic regime is said \emph{rough}.}  In this context, the reference situation is a granular bed composed of natural sand grains of size $d$, which leads to a hydrodynamic roughness $z_0=r d$ in the rough regime. For a static sand bed of grain size $d$, the hydrodynamic roughness measured by extrapolation of the velocity log-profile (\ref{LogProfile}) in the rough regime gives typical values between $r \simeq 0.03 - 0.1$ \citep{bagnold1941physics,kamphuis1974determination,schlichting2000boundary,andreotti2004two}. For an arbitrary rough substrate it has been proposed to defined an \textit{equivalent sand grain} such that its hydrodynamic roughness is the same as that of a sand bed composed of grains of that size \citep{flack2010review,chung2021predicting, kadivar2021review}. This approach assumes that the hydrodynamic roughness is essentially proportional to the average roughness element height \citep{fang1992aerodynamic}.  {Here, we will show, however, that the hydrodynamic roughness presents more subtle dependences on other parameters such as the element spacing density and shape \citep{wiberg1992unidirectional,xian2002field,dong2002aerodynamic,jimenez2004turbulent,cheng2007flow,brown2008wind}.}

\subsection{Governing equations and turbulent closure}
In a spirit similar to \citet{taylor1989parameterization}, we consider that the  {inner surface} roughness is characterised by an equivalent grain size $d$. The smooth-rough transition is controlled by the grain size based Reynolds number $\mathcal{R}_d = d u_*/\nu$. The velocity $\vec{u}$ and pressure $p$ fields are described by means of Reynolds averaged Navier-Stokes equations, which  {states}
\begin{eqnarray}
\partial_i u_i &=& 0\,,
\label{NScont}\\
\rho u_i \partial_i u_j&=&-\partial_j p+\partial_i \tau_{ij}\,, 
\label{NSmomentum}
\end{eqnarray}
where $\rho$ is the fluid density and $\tau_{ij}$ is the Reynolds stress tensor. As a classical mixing theory for boundary layer flows, we take here a first order closure \citep{pope2000turbulent} to relate the stress to the strain rate $\dot \gamma_{ij} = \partial_i u_j+\partial_j u_i$, which adds microscopic molecular and turbulent viscosities:
\begin{equation}
\tau_{ij} = \rho (\nu + \ell^2 |\dot \gamma|) \dot \gamma_{ij} - \frac{1}{3} \rho\chi^2 \ell^2 |\dot \gamma|^2 \delta_{ij}\,,
\label{StrainStress}
\end{equation}
 {where $|\dot \gamma| = \left( \tfrac{1}{2} \dot \gamma_{ij} \dot \gamma_{ij} \right)^{1/2}$ is the mixing frequency proportional to the modulus of the strain rate.}
$\chi$ is a phenomenological constant, which is in the range $2$--$3$ \citep{pope2000turbulent}. It has no importance here as only the normal stress difference $\tau_{xx}-\tau_{zz}$ matters in the calculations \citep{fourriere2010bedforms, claudin2017dissolution}. $\ell$ is the mixing length, which is the key input  {of the model. The simplest expression of the mixing length presenting a smooth-rough transition is $\ell=\kappa (z+rd)$, where the grain-induced roughness $rd$ is added to the geometrical distance to the solid, $z$. However, this formula overestimates turbulent mixing in the viscous sublayer, in the smooth hydrodynamical regime. A classical empirical approach is to introduce a factor to the geometrical length contribution \citep{van1956turbulent} which exponentially kills turbulent mixing when the local Reynolds number $u_*z/\nu$ is smaller than a transitional value $\mathcal{R}^0_t \simeq 25$  \citep{pope2000turbulent}.} We adopt here an expression introduced and tuned in the context of the linear stability analysis for sedimentary or dissolution bedforms \citep{richards1980formation, ayotte1994impact, colombini2004revisiting, fourriere2010bedforms, charru2013sand, claudin2017dissolution}
\begin{equation}
\ell=\kappa (z+rd-Z) \left[1-\exp\left(-\frac{(\tau_{xz}/\rho)^{1/2}(z+sd-Z)}{\nu \mathcal{R}_t}\right)\right].
\label{ellcombo}
\end{equation}
$\kappa$ is the von K\'arm\'an constant as in Eq. \ref{LogProfile}. The parameter $s$ controls the induction of turbulent fluctuations in the viscous layer upon increasing the surface roughness. Both $r \simeq 1/30$ and $s \simeq 1/3$ have been calibrated using measurements of velocity profiles over varied rough walls \citep{schultz2009turbulent, flack2010review} (figure~\ref{schematic}b). In the smooth regime ($\mathcal{R}_d \ll 1$), these numbers give a hydrodynamic roughness controlled by the viscous length: $z_0 \simeq \nu/(7 u_*)$.  {In the rough regime ($\mathcal{R}_d \gg 1$), it rather gives $z_0 \simeq d/30$.}

 {
The model needs to account for the hydrodynamic response of the flow to the perturbation generated by the surface elevation profile, in  the van Driest exponential term. Let us call $\lambda$ the typical horizontal length scale of this profile -- $\lambda$ will later be the wavelength of a particular Fourier mode. When $\lambda$ is  small compared to the viscous sub-layer thickness, the flow perturbation induced by this topography stays confined in the viscous layer. In contrast, when $\lambda$ is much larger than $\nu/u_*$, the perturbation penetrates the outer turbulent zone so that the flow response is turbulent. In the transitional regime between laminar and turbulent responses, the surface bumps generate turbulent perturbations that develop downstream, due to an adverse pressure gradient (diverging mean streamlines). This effectively modulates the thickness of the viscous sub-layer. In the mixing length expression (\ref{ellcombo}), this translates into a non-constant value of $\mathcal{R}_t$. To describe these variations, \citet{hanratty1981stability} proposed an empirical relaxation equation in which the viscous sub-layer thickness lags behind the pressure gradient by a distance scaling with the viscous length. Here we write it in terms of $\mathcal{R}_t$ as
}
\begin{equation}
\frac{u_*}{\nu}(\mathcal{R}_t-\mathcal{R}_t^0)+a \partial_x \mathcal{R}_t=b\frac{ \mathcal{R}_t^0}{\rho u_*^2} \partial_x (\tau_{xx}-p)\,.
\label{HanrattyRelaxation2}
\end{equation}
Both constants $a \simeq 2000$ and $b \simeq 35$ have been calibrated in \citet{charru2013sand} on measurements of the basal shear stress over a modulated surface \citep{zilker1977influence,zilker1979influence,frederick1988velocity}.  {The predicted amplitude of the flow response to topography is very sensitive to these values. In the smooth case, this response presents strong variations for a narrow range of $\lambda$, at the transition from a viscous response to a turbulent response  \citep{claudin2017dissolution}}. In the rough case, this transitional regime disappears as the surface roughness destabilises the viscous sub-layer. We will show that this hydrodynamic `anomaly' associated with the laminar-turbulent transition, has a signature in the outer hydrodynamic roughness.

\subsection{Weakly non-linear expansion}
 {We consider a solid surface of vertical elevation profile $Z(x)$, which is decomposed in Fourier modes. The computation of the outer hydrodynamic roughness induced by $Z(x)$ from the flow response to this topographical perturbation is a non-linear calculation, as one needs to compute the correction to the base (uniform) flow. We perform this calculation below, starting from the base flow and then expanding the governing equations at second order in amplitude of the $Z$ variations, considering only their influence on the zero mode, i.e. on the average flow}. Without loss of generality, a single Fourier mode of wavenumber $k=2\pi/\lambda$ and amplitude $\zeta$ can be considered:
\begin{equation}
Z(x)=\zeta e^{ikx}.
\label{BedProfile}
\end{equation}
$\zeta$ is the amplitude of the surface modulation, and the following perturbation theory will be developed in powers of $k\zeta$ assumed small. We use standard notations with complex numbers for all quantities involved  {for} the linear development in the sake of mathematical convenience, though only real parts are understood. We introduce the dimensionless vertical coordinate $\eta = k z$,  the wavenumber-based Reynolds number $\mathcal{R} = {u_*}/{k \nu}$, the rescaled equivalent grain diameter $\eta_d = kd=\mathcal{R}_d/\mathcal{R}$, and the dimensionless mixing length $\Upsilon = k\ell$.

The base state corresponds to a homogenous substrate in the $x$-direction ($\zeta=0$). In this case, the strain rate reduces into $\partial_z u_x$, and correspondingly Eq.~\ref{NSmomentum} reduces to $\partial_z \tau_{xz}=0$, or equivalently $\tau_{xz}  \equiv \rho |u_*|u_*$ once integrated. The mixing length then  {simplifies into}:
\begin{equation}
\Upsilon = \kappa (\eta+r \eta_d) \left[1-\exp\left(-\frac{\mathcal{R}(\eta+s\eta_d)}{\mathcal{R}_t^0}\right)\right].
\label{UpsilonBaseState}
\end{equation}
Summing up the turbulent and viscous contributions of the shear stress $\tau_{xz}$, (\ref{StrainStress})  {yields}:
\begin{equation}
\rho \left( \ell^2 |\partial_z u_x| \partial_z u_x + \nu \partial_z u_x \right) = \rho |u_*|u_*\,.
\label{ConstantShear}
\end{equation}
We define the function $\mathcal{U}(\eta)$ giving the flow velocity profile in the base state as $u_x \equiv u_* \mathcal{U}$. Following the above equation, $\mathcal{U}$ obeys:
\begin{equation}
\Upsilon^2 |\mathcal{U}'|\mathcal{U}'+ {\mathcal{R}}^{-1} \mathcal{U}'=1,
\qquad {\rm or\ equivalently} \qquad
\mathcal{U}'=\frac{-1 +\sqrt{1+4\Upsilon^2 {\mathcal{R}}^2}}{2\Upsilon^2 {\mathcal{R}}}\,,
\label{EqDiffmathcalU}
\end{equation}
 {where primed} quantities denote derivatives with respect to $\eta$. This equation is solved with the boundary condition $\mathcal{U}(0)=0$ corresponding to the no-slip condition of the fluid at the solid surface.  The viscous sublayer is described by $\mathcal{U}' = \mathcal{R}$ in the limit of negligible $\Upsilon^2 {\mathcal{R}}^2$. Figure~\ref{schematic}b shows a typical velocity profile of a base state, with the characteristic outer logarithmic dependence in $z$ from which the hydrodynamic roughness $z_0$ is extracted. Data available in the literature are collected to demonstrate that the predicted velocity profiles by the present model agree well with experimental measurements.

 {Next, we seek} to compute the flow in response to the surface perturbation (\ref{BedProfile}) one step further than in previous papers by \citet{jackson1975turbulent}, \citet{sykes1980asymptotic}, \citet{hunt1988turbulent}, \citet{fourriere2010bedforms} and \citet{claudin2017dissolution}. At the linear order, Fourier modes can be treated independently of each other, and the computation of the response to a pure sinusoidal profile (\ref{BedProfile})  {provides} the complete field. At the next quadratic order in $(k\zeta)^2$, the computation rules when taking products for non-linear contributions show that these quadratic terms are of two kinds, those in $e^{2ikx}$ and those homogeneous in $x$. Here, only the latter, which provide corrections to the base flow, matters. For our purpose, we consequently write the weakly nonlinear expansion for $u_x$ as
\begin{equation}
u_x = u_* \left[{\mathcal{U}}+(k\zeta) e^{ikx} U_1 + (k\zeta)^2 U_{0}\right]\,,
\label{VelocityNonLinearExpansion}
\end{equation}
where $U_1$ and $U_0$ are the modal functions respectively for the linear and homogeneous quadratic responses. The systematic expansion of Navier-Stokes equations and of the Reynolds averaged closure around the undisturbed profile ${\mathcal{U}}$ is summarised in Appendix \ref{NonLinearRules}. An important point, beside technical details, is that the validity of this expansion has a limited range of $k\zeta$, which is not known in advance. Within the same expansion framework, different representations of the solution can be built, which have different ranges of validity (see Appendix \ref{ShiftedRepresentationStreamFunction}).  {In general, the elevation profile $Z$ presents different Fourier modes, and their contributions at the linear and homogeneous quadratic orders are additive. Importantly, even if all modes contribute to build the hydrodynamic roughness, they do not interact. To account for the small-scale surface corrugations one could have for example considered them as part of the elevation profile, and worked with two well-separated modes. Here, however, the surface roughness is encoded in the turbulent mixing length (\ref{ellcombo}) by means of two terms expressed with the equivalent sand grain size $d$, and it affects the base velocity profile ${\mathcal{U}}$, the turbulent mixing at the surface and the homogeneous flow response $U_{0}$. One could then wonder whether these two ways are equivalent, and we shall see in the next section that mode coupling is required to obtain self-consistent results and to recover the smooth-rough transition.}\\
\begin{figure}
\centering
\includegraphics[scale=1.0]{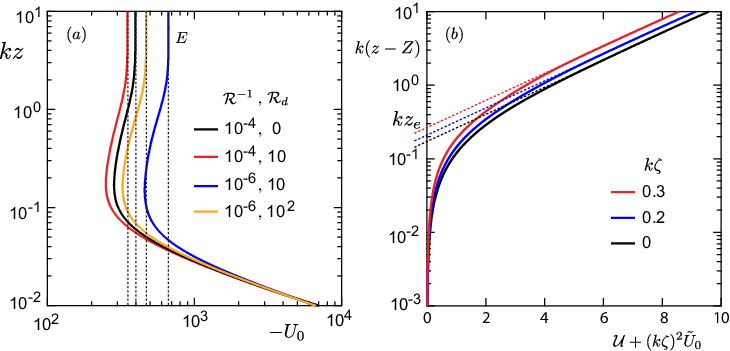}
\caption{(a) Vertical profiles of the homogeneous second order correction to the horizontal velocity for different values of $\mathcal{R}^{-1}$ and $\mathcal{R}_d$ (legend), where $-U_0$ is shown as a function of $\eta = kz$. (b) Homogeneous part of the velocity $\mathcal{U}+(k\zeta)^2\tilde U_0$ as a function of the distance to the surface $k(z-Z)$ for three values of $k\zeta$ (legend) {and for $\mathcal{R}^{-1}=0.05$ and $\mathcal{R}_d = 10$}. The extrapolation to $0$ of the upper part of the profiles gives the values of the effective hydrodynamic roughness $kz_e$ seen far from the surface. 
}
\label{U0_eta_Rd}
\end{figure}

\subsection{Self-consistent looped expansion}
\label{LoopedExpansion}
 {A simple way of coupling the different modes of the elevation profile is to ensure that the base flow used for the expansion incorporates the corrections due to the corrugations of the solid surface at all scales. This is fundamentally different from the introduction of new terms in the mixing length. In practice, we use the following iterative procedure. For each mode of given $k\zeta$ and $\mathcal{R}$, we compute the quadratic correction $U_0$ to the flow induced by this perturbation using the base flow velocity profile $\mathcal{U}$ as described in the previous paragraph. We deduce the corrected homogeneous velocity profile $\hat{\mathcal{U}} = \mathcal{U} + (k \zeta)^2 U_0$ and compute the next correction using $\hat{\mathcal{U}}$ (instead of $\mathcal{U}$) for the new base flow in the weakly non-linear expansion described in Appendix {A}: all convective turbulent mixing terms are then computed with a velocity profile that accounts for that perturbation. We repeat this loop until convergence to a final profile $\hat{\mathcal{U}}$ which, used in the expansion, leads to the profile of $U_0$ giving that $\hat{\mathcal{U}}$. This self-consistent expansion around the disturbed average velocity profile, is {\it a priori} more precise than the non-looped expansion. This approach resembles, in spirit, implicit (as opposed to explicit) schemes in numerical solvers of differential equations. The price to pay, however, is this iterative procedure that must be performed for \emph{each} value of $k \zeta$.}

\section{Results}
\label{TopographyInducedHydrodynamicRoughness}

 {We present in this section the results obtained with the above model to compute the effective hydrodynamic roughness $z_e$ induced by the surface elevation profile (\ref{BedProfile}). We start with the standard perturbative calculation, which will allow us to discuss different regimes and to evidence the signature of the laminar-turbulent transition on $z_e$. With the looped calculation, we will then show how to match inner and outer roughnesses, with application to the smooth-rough transition as well as to superimposed bedforms.}

\subsection{Hydrodynamic roughness from homogeneous quadratic velocity correction}
Following (\ref{VelocityNonLinearExpansion}), the homogeneous part of the velocity profile shows that the base state $\mathcal{U}$ is corrected by the quadratic term $(k\zeta)^2 U_0$. The modal function $U_0$ associated with this correction is  {illustrated} in figure \ref{U0_eta_Rd}a, for various values of $\mathcal{R}$ and $\mathcal{R}_d$. Its vertical profile tends towards a constant $U_0 \sim-E$ far from the surface, approximately for $z \gtrsim \lambda$.  {We can define the outer hydrodynamic roughness from this asymptotic behaviour of the function $U_0$, and $E$ is accordingly referred to as \emph{roughness coefficient}.} Using the fact that the base profile $\mathcal{U} \sim \frac{1}{\kappa}\ln\frac{z}{z_0}$ in that limit of large $z$, and identifying the corresponding behaviour of the corrected velocity profile $\mathcal{U}+(k\zeta)^2 U_0$ with a similar log-profile $\frac{1}{\kappa}\ln\frac{z}{z_e}$ of effective roughness $z_e$, we simply obtain
\begin{equation}
\ln \frac{z_e}{z_0} = \kappa (k \zeta)^2 E.
\label{defzg}
\end{equation}
Figure \ref{U0_eta_Rd}b shows the velocity profiles in the shifted representation (Appendix \ref{ShiftedRepresentationStreamFunction}). One observes that $z_e$ indeed increases with $k\zeta$. Interestingly, the ratio $z_e/z_0$ does not only depend on the amplitude $\zeta$ of the surface elevation, but also on $\mathcal{R}$ and $\mathcal{R}_d$. We describe and analyse below the variations of the coefficient $E$, which encodes the dependence with respect to these two quantities.

\subsection{Roughness coefficient $E$ in the different regimes}
\begin{figure}
\centering
\includegraphics[scale=1]{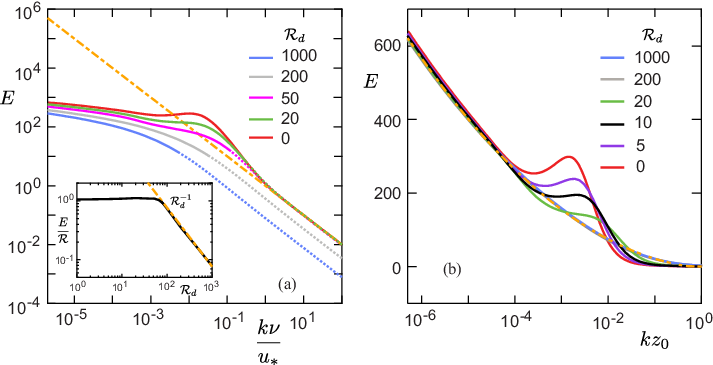}
\caption{ (a) Roughness coefficient $E$ as a function of $k \nu/u_* = \mathcal{R}^{-1}$ for different $\mathcal{R}_d$ (legend). A line collapse is observed in the limit of large wavenumber when $\mathcal{R}_d$ is small, which is well fitted with $E \sim \mathcal{R}$ (dash-dotted orange line). Dashed lines correspond to the moment at which $\lambda$ becomes smaller than $d$. Inset: $E/\mathcal R$ as a function of $\mathcal R_d$ in the viscous limit ($\mathcal R \to 0$). It is well fitted with $E/\mathcal R \sim \mathcal R_d^{-1}$ when $\mathcal R_d \gtrsim 100$ (dash-dotted orange line). (b) The same coefficient $E$ but now as a function of $kz_0$. A line collapse is observed in the limit of small wavenumber. The dash-dotted orange curve is the fit of the rough curves by Eq.~\ref{scalingERoughTurbulent}, with $a \simeq 2.7$, $b \simeq -4.1$ and $c \simeq 0$.}
\label{E}
\end{figure}

The variations of the coefficient $E$ on the parameters $\mathcal{R}$ and $\mathcal{R}_d$ are  {illustrated} in figure~\ref{E}. For a given $\mathcal{R}_d$, $E$ essentially decreases with $k\nu/u_* = \mathcal{R}^{-1}$ (panel a), with $E$ systematically smaller for larger $\mathcal{R}_d$. Below $k\nu/u_* \simeq 10^{-2}$, the variations are rather weak (but note the vertical log-scale). Above $k\nu/u_* \simeq 10^{-1}$, the curves become straight and steeper, corresponding to $E \propto \mathcal{R}$, with a collapse at small $\mathcal{R}_d$. When plotted as a function of $kz_0$, the coefficient $E$ is also  {generally} decreasing, except within an intermediate range $k z_0 \simeq 10^{-4}$--$10^{-1}$, where non-monotonic variations develop, all the larger for smaller $\mathcal{R}_d$ (panel b). Below $k z_0 \simeq 10^{-4}$, a collapse along a parabolic behaviour is observed. This non-monotonic behaviour is in fact also noticeable in panel (a), but less visible due to the vertical log-scale.  {Below, we comment in more detail these different regimes, and provide, when possible, some asymptotic scaling behaviours.}

 {In the viscous regime, the flow perturbation is imbedded in the viscous sub-layer. In this limit $\mathcal{R}$ is small and one can see in figure~\ref{E}a that this regime starts at $k\nu/u_* \gtrsim 10^0$, which roughly corresponds to $kz_0 \gtrsim 10^{-1}$. Since turbulent fluctuations are not relevant in this limit, the mixing length vanishes and the shear stress is dominated by the viscous term $\tau_{xz} \sim \rho \nu \partial_z u_x$. It follows that all stress modal functions are simply related to the derivatives of those of the velocity as: $S_{t1} \sim \mathcal{R}^{-1} U_1'$ and $S_{t0} \sim \mathcal{R}^{-1} U_0'$. The non-linear correction to the stress $S_{t0}$ can only come from the inertial term in the momentum balance of Navier-Stokes equations, which becomes negligible in this limit. As a consequence, $U_0' \to 0$, so that $U_0$ tends to a constant, whose value can be deduced from the lower boundary condition. In order to get no slip at the surface, i.e. $u_x(Z)=0$ at second order in $k\zeta$, $U_0$ must compensate the variation of $U_1$: $-U_0 (0) \sim U_1'(0)$. Relating this derivative to the stress, we then obtain $-U_0 (0) \sim \mathcal{R} S_{t1}(0)$. A proper derivation of the equations in this asymptotic regime (see Appendix~\ref{AsymptoticRegimes}) leads to}
\begin{equation}
E = -U_0 \sim \frac{1}{2} \mathcal{R} \mathcal{A}\,,
\label{ViscousScalingE}
\end{equation}
 {where the basal shear coefficients $S_{t1}(0) = \mathcal{A} + i \mathcal{B}$ have been introduced \citep{fourriere2010bedforms,claudin2017dissolution}}. This asymptotic scaling with respect to $\mathcal{R}$ is nicely verified by the full model (orange line in figure~\ref{E}a). Furthermore, all curves collapse in this viscous limit for the smaller values of $\mathcal{R}_d$. This is consistent with the fact that the in-phase stress coefficient $\mathcal{A} \to 2$ in the smooth regime \citep{benjamin1959shearing, charru2000phase, charru2013sand}, so that one expects that $E \approx \mathcal{R}$, which is indeed what we obtain up to $\mathcal{R}_d \simeq 100$.

For larger values of $\mathcal{R}_d$, the curves in figure~\ref{E}a at large wavenumber decollapse but  {remains} parallel in this log-log representation. This means that they share the same dependence $\propto \mathcal{R}$, but with a prefactor that decreases with $\mathcal{R}_d$. Having both small $\mathcal{R}$ and large $\mathcal{R}_d$  {may appear to be} an unphysical limit, as it would correspond to a surface perturbation at a scale $\lambda$ smaller than the equivalent roughness size $d$. However, one could for example imagine non-geometrical sources of effective surface roughness, such  {as} random jet injection \citep{park1999effects,kametani2015effect}, that could  {obey} this scale hierarchy. In this case, a simple argument is to keep the scaling $E \sim \mathcal{R}$, but replacing the viscosity involved in the definition of $\mathcal{R}$ by an effective turbulent viscosity $\nu_t \sim u_* d$, such that 
\begin{equation}
E \sim  \frac{u_*}{k \nu_t} \sim \mathcal{R}/{\mathcal{R}_d}\,.
\label{ViscousTurbulentScalingE}
\end{equation}
This decreasing behaviour in $1/\mathcal{R}_d$ at large $\mathcal{R}_d$ is  nicely verified by the dashed part of the curves in figure~\ref{E}a (inset).

Let us turn now to the fully turbulent regime, typically for $k\nu/u_* \lesssim 10^{-3}$, corresponding to $kz_0 \lesssim 10^{-4}$. In this limit, one can infer an expression for $E(kz_0)$ estimating $U_0$ at a typical (dimensionless) height $\eta \sim 1$, above which its profile becomes constant (figure\ref{U0_eta_Rd}a). For that purpose, we shall combine estimations of the shear stress from the momentum balance (\ref{NSmomentum}) and from the turbulent closure  (\ref{StrainStress}).

 {
The momentum balance essentially writes $\rho u \partial_x u \sim \partial_z \tau$. Because the quadratic response we are interested in is homogeneous in $x$, the only way to contribute to the homogeneous stress correction is $U_1^2 \sim S_{t0}$, where we have also used the fact that horizontal and vertical derivatives both scale as $\partial_x \sim \partial_z \sim 1/\eta \sim 1$. Similarly, the relation between stress and strain rate (\ref{StrainStress}) can be expressed in a scaling way as $\tau \sim \rho \ell^2 \left( \nabla u \right)^2$. For the homogeneous correction in $(k\zeta)^2$, the LHS of this relation is again $S_{t0}$, but we expect for its RHS several contributions from the product of the different factors, combining their expressions in the base state, at linear order $1$ and quadratic homogeneous order $0$. Gathering these contributions, and as more detailed in Appendix~\ref{AsymptoticRegimes}, one essentially obtains $S_{t0} \sim U_0 + U_1^2 + U_1 + \mbox{Cst}$. Finally, using the relation for the asymptotic expression  {for} the linear correction of the velocity \citep{fourriere2010bedforms} $U_1 \sim (\mathcal{A}+i\mathcal{B})/(2\kappa) \ln (\eta/k z_0) \sim \ln (k z_0)$, we finally obtain for $E \sim -U_0$ at $\eta \sim 1$:}
\begin{eqnarray}
E \simeq a \ln^2 kz_0 + b \ln kz_0 +c.
\label{scalingERoughTurbulent}
\end{eqnarray}
The coefficients $a$, $b$ and $c$  {are expected to} be approximately constant, since they come from $\mathcal A$ and $\mathcal B$, which are only weakly dependent on $kz_0$ in the turbulent regime \citep{fourriere2010bedforms,charru2013sand}. Their adjustment to the result of the full integrated model gives a good fit  { to the data for hydrodynamically rough conditions} (figure~\ref{E}b), in practice valid on the entire range of $kz_0$. Although not directly expressed in terms of such a coefficient $E$, a similar quadratic behaviour was derived by \citet{jacobs1989effective} by means of asymptotic analysis of RANS equations closed by Launder-Spalding turbulence model. \citet{taylor1989parameterization} also report linear variations of the equivalent of $E$ with $\ln kz_0$ in the range $10^{-8}$--$10^{-3}$, with different turbulent closures.

Finally, at intermediate wavenumbers, one observes a non-monotonic variation of $E$ with respect to $kz_0$. It  significantly increases the effective roughness around $kz_0 \simeq 10^{-3}$. It is most pronounced when the flow is hydrodynamically smooth and it becomes less and less  {significant as} $\mathcal{R}_d$ increases (figure \ref{E}b). This striking behaviour is similar to what is also observed in the flow linear response for the behaviour of the stress coefficients $\mathcal A$ and $\mathcal B$ \citep{charru2013sand, claudin2017dissolution}, and the pressure coefficients $\mathcal C$ and $\mathcal D$ \citep{claudin2021basal}, when $10^{-4} \lesssim kz_0 \lesssim 10^{-2}$.  {This `anomaly' takes place at the transition between a laminar and a turbulent response of the flow to the elevation profile, discussed above. It is included in the description by the relaxation equation of the transitional Reynolds number $\mathcal{R}_t$ (Eq.~\ref{HanrattyRelaxation2}), associated with this laminar-turbulent transition. A qualitative picture of this transition is that turbulent bursts are produced at the crest and develop on the downstream side \citep{charru2013sand}, when the disturbance reaches the size of the viscous boundary layer. This effect shares some similarities with the drag crisis on a sphere or a circular cylinder \citep{choi2008control}.} A deeper understanding of it is however missing and new experiments as well as numerical simulations are definitively needed to go further on this fundamental question.

\subsection{Matching between inner and outer scales}
\label{Matching}
 {
So far, we have modelled the inner roughness by terms in the mixing length proportional to the equivalent grain size $d$. Consider again the case of grains at the surface of sand ripples superimposed to a large dune. At the largest scale, the sand ripples are the source of dune-surface roughness. They then could be also treated by a modified mixing length with a corresponding equivalent grain size. However, one can zoom on these sand ripples and consider them as geometrical disturbances to the dune surface, rather than a source of turbulent mixing. Similarly, zooming on the grains, one expects the roughness they induce to be an emergent property of their geometrical effect. The model should therefore be self-consistent when considering the induced roughness from the two perspectives: a source of turbulent mixing and a geometrical corrugation.
}

\begin{figure}
\centering
\includegraphics[scale=1.0]{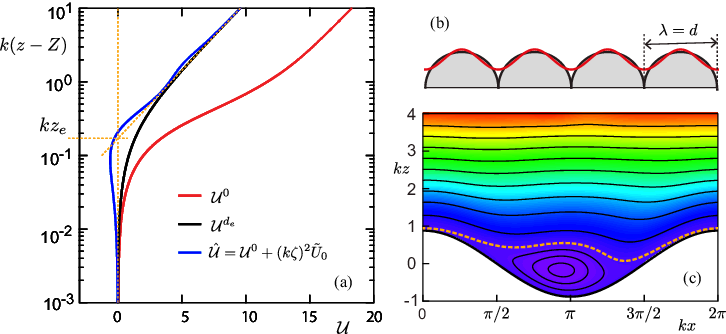}
\caption{(a) Velocity profiles $\mathcal{U}^0$, $\hat{\mathcal{U}}$ and $\mathcal{U}^{d_e}$ introduced in the self-consistent calculation. These two last functions overlap in the upper part of their profiles, i.e. possess the same hydrodynamic roughness $z_e$ (dashed orange lines). Here computation done for $k\zeta =0.88$ (see figure~\ref{kz0_Rdfit2}a) and $k\nu/u_* = 0.05$, which converges to $d_e u_*/\nu \simeq 130$.
(b) Sinusoidal approximation of the surface of a periodic array of touching grains, setting $\lambda = d$. 
(c) Isocontours of the dimensionless stream function $k \Psi /u_*$ (see Appendix \ref{ShiftedRepresentationStreamFunction} ) above a modulated surface (in white), when $k\zeta =0.88$ and $k\nu/u_* = 0.05$. The flow is from left to right. Color code: $k \Psi /u_*$ is increasing from blue to red in the background; and the dashed orange curve corresponds to $k \Psi /u_* = 0$.}
\label{kz0_Rdfit}
\end{figure}

 {As a first test, we consider a flat rough surface whose corrugations are modelled by a smooth surface of wavelength $\lambda$, which will be set by the equivalent grain size $d$ (figure~\ref{kz0_Rdfit}b). However, to avoid confusion, we will keep the notation $\lambda$. Following the perturbative prediction (\ref{defzg}) for a smooth surface, the effective hydrodynamic roughness induced by such an elevation of amplitude $\zeta$ reads
\begin{equation}
z_e \simeq \frac{\nu}{7 u_*} \exp \left[ \kappa (k \zeta)^2 E \right],
\label{zedirect}
\end{equation}
where $E$ is given by the red curve in figure~\ref{E}a (smooth limit), computed for $k = 2\pi /d$. When plotted as a function of $d u_*/\nu$ to compare with the data (dashed line in figure~\ref{schematic}a, for $k\zeta =0.88$, see below), we see that the agreement is good in the smooth regime, but starts to diverge from the measurements when the roughness becomes on the order of the viscous sublayer: $z_e u_*/\nu \simeq 1$. This expansion is thus unable to reproduce the transition towards the rough regime, and predicts a hydrodynamic roughness significantly larger than that observed: the dependence is exponential rather than linear. The problem lies in the fact that, when the surface corrugation increases, it induces inertial mixing of momentum that destroys the viscous sublayer. The smooth base profile with its viscous sublayer is thus not the relevant uniform state around which the expansion should be performed.
}

 {
The idea is then to use instead the looped expansion introduced in section~\ref{LoopedExpansion}. We start with the the smooth base velocity profile $\mathcal{U}^0$, i.e. computed from (Eq.~\ref{EqDiffmathcalU}) with $\eta_d=0$ in (\ref{UpsilonBaseState}), choose a value of $k\zeta$, and iterate the described procedure until convergence. To avoid numerical instabilities, we use the numerical trick to approximate at each loop $\hat{\mathcal{U}}$ by the base profile $\mathcal{U}^{d_e}$ computed for a non-vanishing $d_e$, so that their hydrodynamic roughnesses are the same (figure~\ref{kz0_Rdfit}a).
}

\begin{figure}
\centering
\includegraphics[scale=1.0]{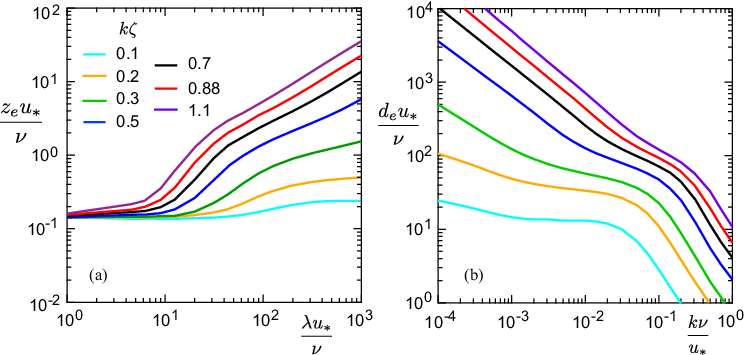}
\caption{ (a) Dimensionless effective hydrodynamic roughness $z_e u_*/\nu$ as a function of $\lambda u_*/\nu$, computed in a self-consistent way for different values of $k\zeta$ (see legend). (b) Resulting equivalent grain size $d_e u_*/\nu$ as a function of $k \nu /u_*$ for different $k\zeta$ (same colour code as in panel a).}
\label{kz0_Rdfit2}
\end{figure}

We show in figure~\ref{kz0_Rdfit2}a the effective roughness computed in this self-consistent manner for various values of $k\zeta$, as a function of $\lambda u_*/\nu$.  {It can be seen} that the transition from smooth to rough regimes is indeed recovered: the roughness changes from $z_e u_*/\nu \simeq 1/7$ to a linear increase with $\lambda$. This asymptotic increase is steeper for larger values of $k \zeta$. When compared to the data on rough flat surfaces by setting $\lambda=d$, this prediction fits well the measurements when $k \zeta \simeq 0.88$ (see the red solid curve in figure~\ref{schematic}a). Interestingly, this value corresponds to the sinusoidal approximation of the surface of a periodic array of touching grains $Z \simeq 0.14 \, d \cos \left(2 \pi x/d \right)$, which gives $k \zeta \simeq  0.14 \times 2\pi \simeq 0.88$ (red line in figure~\ref{kz0_Rdfit}b).

Although the direct and self-consistent derivations of the effective hydrodynamic roughness (dash-dotted and solid red curves in figure~\ref{schematic}a, respectively) are both expansions at the second order in aspect ratio $k\zeta$, their range of validity is rather different. The better performance of the self-consistent approach is due to the fact that it takes into account the topography-induced velocity reduction at the scale of $\lambda$. Consistently, we can interpret the observation that the self-consistent derivation predicts a lower $z_e$ than the direct calculation, as accounting for flow recirculations at the scale of the topography reduces the effective corrugation of the bottom (figure~\ref{kz0_Rdfit}c).

\begin{figure}
\centering
\includegraphics[scale=1.0]{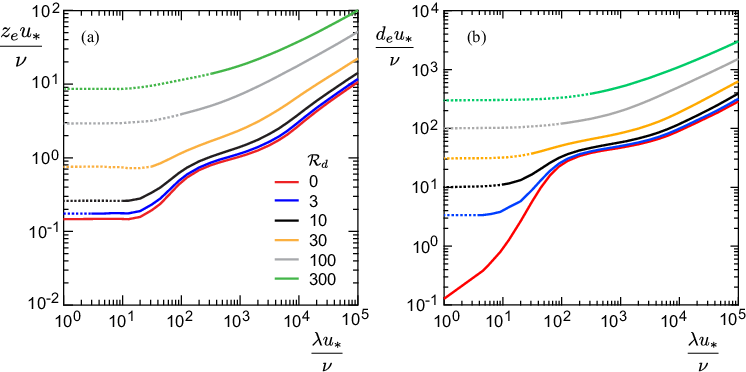}
\caption{Self-consistent calculation of the effective roughness \emph{vs} bedform wavelength (a) and the equivalent grain size \emph{vs} bedform wavelength (b). These are computed for a perturbation of various surface roughness $\mathcal{R}_d$ encoded by different colours (see legend for the colour code in both panels) and with an aspect ratio $k\zeta \simeq 0.25$, typically representative of sedimentary dunes. Dashed lines correspond to the moment at which $\lambda$ becomes smaller than $d$.
}
\label{Superimposedbedform}
\end{figure}

 {
\subsection{Superimposed sedimentary bedforms}
\label{SuperimposedBedforms}
The other situation where inner and outer roughnesses must match is that of superimposed beforms: the outer roughness of a given level in the scale hierarchy is the inner roughness of the next one, as we illustrated with grains, ripples and dunes (figure \ref{multiscale}a). For this purpose, the looped calculation described in section~\ref{LoopedExpansion} must be initiated at some finite value of $\mathcal{R}_d$.} The resulting curves are displayed in figure~\ref{Superimposedbedform} for $k \zeta = 0.25$, a typical value for dunes at equilibrium (aspect ratio on the order of $1/12$). It shows consistent increasing trends for larger and rougher bed. The curves start to diverge from the smooth case when $\mathcal{R}_d$ is larger than a few units. In addition, the resulting effective roughness can be further transformed into the Darcy friction factor (or skin friction) and the hydraulic resistance (or form drag) of bedforms \citep{engelund1977hydraulic, zanke2022roughness}.  {An empirical formula for the hydrodynamic roughness of sedimentary bedforms has been proposed by \citet{van1984sediment}, which reflects the growth of the hydrodynamic roughness from a value selected by the grain diameter $d$ to a value selected by the bedform wavelength, as the aspect ratio $k\zeta$ increases. At small aspect ratio, for $k\zeta \lesssim 0.25$, which includes the typical aspect ratio of sedimentary ripples and dunes, the looped calculation performed here turns out to be recovering values close to this empirical prediction.}

This situation of multi-scaled pattern is particularly relevant in the sub-aquatic case, as well as for low-pressure -- e.g. planetary -- gaseous environments, which provide viscous or transitional conditions \citep{colombini2011ripple, jia2017giant, duran2019unified, andreotti2021lower, gunn2022conditions}. We shall illustrate here how to apply the present results on the emblematic example of Martian dunes and ripples (figure~\ref{multiscale}b,c). The dunes, whose typical length is on the order of a few hundreds of meters, exhibit large meter-scale ripples on their surface; these ripples themselves are superimposed with decimeter-scale small ripples \citep{lapotre2016large, lapotre2018morphologic}. A hydrodynamic-based explanation for the existence of these intermediate bedforms has been proposed by \citet{duran2019unified},  {and verified by \citet{rubanenko2022distinct}:} the anomaly associated with the laminar-turbulent transition creates a forbidden gap in wavelength that stops the coarsening of the most unstable mode. As discussed above, this anomaly is present in the smooth regime only. Estimating Martian conditions with a grain size $d \simeq 100~\mu$m, a shear velocity $u_* \simeq 0.5$~m/s and a viscosity $\nu \simeq 10^{-3}$~m$^2$/s, one obtains $\mathcal{R}_d \simeq 0.05$, i.e. a hydrodynamically smooth granular bed. The question is then to evaluate, for the large meter-scale ripples, the hydrodynamic roughness induced by their superimposed small ripples. Taking, for these small ripples, a wavelength $\lambda \simeq 0.1$~m and an aspect ratio on the order of $1/20$, i.e. $k\zeta \simeq 0.15$, one obtains, with $\lambda u_*/\nu \simeq 50$, an effective roughness $z_e u_*/\nu \simeq 0.2$ corresponding to an equivalent sand grain Reynolds number $\mathcal{R}_{d_e} \simeq 5$. This indicates that the flow over the large ripples is indeed in the smooth regime, consistently with the hypothesis of \citet{duran2019unified}. Now taking for these large ripples $\lambda \simeq 3$~m, i.e. $\lambda u_*/\nu \simeq 1.5 \times 10^3$, and $k\zeta \simeq 0.25$, we see in figure~\ref{Superimposedbedform} ($\mathcal{R}_{d} = 5$ is between the blue and black lines) an effective roughness $z_e u_*/\nu \simeq 1.5$, corresponding to an equivalent sand grain Reynolds number $\mathcal{R}_{d_e} \simeq 55$. This means that the flow over the large dunes should be considered intermediate between smooth and rough regimes.  {Importantly, aeolian saltation -- and sediment transport in general \citep{van1984sediment} -- is known to increase the sediment bed roughness, with $z_0$ on the order of a few $d$ in the terrestrial case, associated with Bagnold's focal point \citep{duran2011aeolian,valance2015physics}. In low-pressure (Martian) conditions, no such measurement is available so far, but we have checked that the above conclusions are unchanged when doing this hierarchical looped calculation with an initial similar saltation-induced surface roughness.}

 {
\subsection{Non-linear size selection of dissolution bedforms}}
\label{DissolutionBedforms}
Finally, the self-consistent prediction of the effective roughness induced by a wavy surface can be used to revisit the argument given by \citet{claudin2017dissolution} for the selection of the aspect ratio of dissolution patterns. The instability at the origin of these bedforms is associated with the hydrodynamic anomaly on smooth modulated beds for a narrow range of wavenumbers around $k\nu/u_* \simeq 10^{-3}$. This anomaly, and thus the instability mechanism, disappears when the bed is rough enough. Because the development of the bedforms generates an effective roughness, the idea is that they would stop growing when this threshold in $\mathcal{R}_d$ is reached. Here we can make this argument more quantitative with the proposed non-linear calculations. The looped computation of $z_e$ equivalently provides a corresponding equivalent grain size $d_e$, which depends on $k\zeta$ (figure~\ref{kz0_Rdfit2}b). Experimental studies on the development of ice ripples and scallops report equilibrium shapes with the  {inverse} Reynolds numbers in the range  {$k\nu/u_* \simeq 8 \times 10^{-4}$\, -- \,$2 \times 10^{-3}$}, and aspect ratios $6$--$8$ \%, i.e. $k\zeta \simeq 0.19$--$0.25$ \citep{ashton1972ripples, bushuk2019value}. The corresponding induced roughness reads from figure~\ref{kz0_Rdfit2}b around $d_e u_*/\nu \simeq 50$. This value is nicely consistent with the instability threshold computed for parameters relevant to the case of ice melting \citep{claudin2017dissolution}.

\section{Concluding remarks}
\label{conclusion}

Pursuing our goal to describe the hydrodynamic response to a bed perturbation, we have here gone one step beyond the linear order, and computed the homogeneous correction to the base flow induced by the topography. This is a quadratic effect in the amplitude of the bed elevation, thus requiring weakly non-linear calculations. Three main results can be emphasised: we quantify how the bed corrugation increases the effective hydrodynamic roughness; we show that this effect is sensitive to the laminar-turbulent transition; this model is able to reproduce the smooth-rough transition in a self-consistent manner and can be applied to multiscale elevation profiles where inner and outer roughnesses are hierarchically nested.

The expression (\ref{defzg}) shows that the effective roughness $z_e$ cannot be simply related to a single geometrical length \citep{schlichting2000boundary,van1982equivalent,raupach1991rough,wiberg1992unidirectional}, here the amplitude $\zeta$ of the surface elevation profile. It still depends on the inner roughness of the flat surface $z_0$. Moreover, it involves a coefficient $E$ that also depends on the wavenumber of the surface profile, typically decreasing with larger $k$. Interestingly, it shows a non-monotonic behaviour for a range of wavenumber. This anomaly is larger for smoother surfaces, and disappears in the rough limit. Consistently with a similar anomaly for the stress and pressure response \citep{claudin2017dissolution,claudin2021basal}, it can be associated, in the model, to the lag between the thickness of the viscous sub-layer (parametrised by $\mathcal{R}_t$) and the pressure gradient (Eq.~\ref{HanrattyRelaxation2}). This anomaly plays a key role in the formation of sedimentary, dissolution or sublimation bedforms, especially in low-pressure planetary conditions \citep{duran2019unified, bordiec2020sublimation}. However, its direct experimental evidence is limited to a series of measurements by Hanratty and co-workers \citep{zilker1977influence,frederick1988velocity}, that involve a dedicated apparatus with electro-kinetic probes to obtain the basal shear stress response. The present work suggests that the measurement of the hydrodynamic roughness induced by a sinusoidal smooth bottom, achievable by more standard and non-intrusive velocimetric techniques, could offer an alternative way to provide experimental evidence for this anomaly and allow for a better calibration of the relaxation equation (\ref{HanrattyRelaxation2}). Importantly, as a larger surface perturbations induce the smooth to rough transition, the self-consistent calculation of $z_e$ shows that the anomaly disappears when $k\zeta$ is too large, and the non-monotonic behaviour of the roughness is effectively observed here typically for $k\zeta \lesssim 0.05$. Finally, these results also shed light on open hydrodynamic topics associated with such transitional shear flows over a solid wall \citep{tuckerman2020patterns,gome2022extreme}.

Numerical simulations provide another way to investigate these questions. In particular, DNS over wavy surfaces, in the spirit of those of \citet{maass1996direct} and \citet{de1997direct}, would allow one to gain a deeper understanding of Hanratty's anomaly. In such simulations, the good control of the imposed flow as well as the surface properties (wavelength, amplitude, roughness), together with a scale separation which requires a resolved viscous sub-layer much smaller than the surface wavelength, itself sufficiently smaller than the system size, is definitively challenging at relevant values of the Reynolds number, but probably reachable with current numerical techniques and computer power \citep{lee2015direct}. The understanding and the description of the interplay between a wavy surface and the associated modulation of the viscous sublayer then remains an interesting open problem both from numerical and experimental points of view, with significant potential applications for geomorphology and geophysical flows.

\appendix

\section{Weakly non-linear expansion}
\label{NonLinearRules}
In this appendix, we provide detailed information on how to perform the weakly nonlinear expansion. The aim is to derive two sets of closed equations for the linear and homogeneous quadratic responses, which are then solved with boundary conditions at the bottom and top of the computing domain.
 
The weakly nonlinear expansion for $u_x$ has been given in Eq. \ref{VelocityNonLinearExpansion} as follows
\begin{equation}
u_x = u_* \left[{\mathcal{U}}+(k\zeta) e^{ikx} U_1 + (k\zeta)^2 U_{0}\right]\,,
\end{equation}
with $U_1$ and $U_0$ being the modal functions respectively for the linear and homogeneous quadratic responses. Herein, standard notations with complex numbers are used for all quantities involved  {for} the linear development in the sake of mathematical convenience, though only real parts are understood. However, when computing the non-linear terms, one must go back to real notations  {through} the transform $f \to (f+f^*)/2$, with $f^*$ being the complex conjugate of $f$. In the same way, expansions are respectively performed for the vertical velocity $u_z $, the shear stress $\tau_{xz}$, the normal stress difference $\tau_{zz} - \tau_{xx} $, the vertical component of the stress $p-\tau_{zz} $ and the dimensionless mixing length $k\ell$ as
\begin{eqnarray}
u_z  &=&  u_* \left[(k\zeta) e^{ikx} W_1\right]\,,\label{W1expansion}
\\
\tau_{xz}&=&  \rho u_*^2 \left[1+(k\zeta) e^{ikx} S_{t1}+(k\zeta)^2 S_{t0}\right]\,,\\
\label{Stexpansion}
\tau_{zz} - \tau_{xx} &=&    \rho u_*^2 \left[(k\zeta) e^{ikx} S_{d1}+(k\zeta)^2 S_{d0}\right]\,,\\
\label{Sdexpansion}
p-\tau_{zz} &=&    \rho u_*^2 \left[\frac{1}{3}\chi^2+(k\zeta) e^{ikx} S_{n1}+(k\zeta)^2 S_{n0}\right]\,,\\
\label{Snexpansion}
k\ell  &=&  \Upsilon +  k\zeta e^{ikx} L_1+(k\zeta)^2 L_{0}\,.
\label{ellexpansion}
\end{eqnarray}
Note that as there cannot be fluid flow through the solid boundary, the term $W_0(\eta)=0$ is not included into the expansion.

With these notations, the strain rate modulus  {can be rewritten as}
\begin{eqnarray}
|\dot \gamma|&=&\mathcal{U}'+(k\zeta) e^{ikx} (U_1'+iW_1)+(k\zeta)^2 \left[U_0'+\frac{1}{\mathcal{U}'}U_1U_1^*\right]\,,
\end{eqnarray}
and following Eq. \ref{StrainStress}, the stress modal functions can be expressed as
\begin{eqnarray}
S_{t1}&=& \left(2 \Upsilon^2
   \mathcal{U}'+\mathcal{R}^{-1}\right) (U_1'+i W_1)+2 L_1 \Upsilon \mathcal{U}'^2\,,
   \label{FunctionSt1}
\\
S_{t0}&=&\left(2 \Upsilon^2 \mathcal{U}'+\mathcal{R}^{-1}\right)U_0'+\frac{1}{2} L_1 \mathcal{U}'^2 L_1^*+ \Upsilon^2
   \left( U_1 U_1^*+\frac{1}{2} (U_1'+i W_1) \left(U_1'^*-i W_1^*\right)\right)\,,\nonumber\\
&+&\Upsilon \left( \mathcal{U}' L_1^* (U_1'+i W_1)+ L_1 \mathcal{U}' \left(U_1'^*-i
   W_1^*\right)+2 L_0 \mathcal{U}'^2\right)\,,
    \label{FunctionSt0}
   \\
   S_{d1}&=& -\frac{4 i}{ \mathcal{U}' }  U_1\,,
   \label{Sd1}
\\
S_{d0}&=&-2i \Upsilon \left( \mathcal{U}' U_1 L_1^*- L_1 \mathcal{U}' U_1^*\right)-i \Upsilon^2 \left(U_1 \left(U_1'^*-i W_1^*\right)-U_1^*(U_1'
   +i W_1 )\right)\,,
\end{eqnarray}
where $(\mathcal{R}^{-1} + \Upsilon^2 \mathcal{U}' ) = 1/ \mathcal{U}'$ as deduced from Eq. \ref{EqDiffmathcalU}. Further, the modal functions for the mixing length are obtained as 
\begin{eqnarray}
\frac{L_1}{\kappa}&=&\frac{1}{2} e^{-(\eta+s\eta_d) \frac{\mathcal{R}}{\mathcal{R}_t^0}} \left\{ (\eta+r\eta_d) \frac{\mathcal{R}}{\mathcal{R}_t^0} \Big[ 2 \left(\eta+s\eta_d \right)R_{t1} +\left(\eta+s\eta_d \right) S_{t1}-2 \Big] +2 \right\}-1\,,\\
\frac{L_0}{\kappa}&=&-\frac{1}{16} \frac{\mathcal{R}}{\mathcal{R}_t^0} \, e^{-(\eta+s\eta_d) \frac{\mathcal{R}}{\mathcal{R}_t^0}}  \left\{ 4 \Big[ (\eta+r\eta_d) \left(\frac{\mathcal{R}}{\mathcal{R}_t^0} -  2 S_{t0} (\eta+s\eta_d)\right) -2 \right. \nonumber \\
&+&\left(R_{t1}^*+R_{t1} +(S_{t1} + S_{t1}^*)/2\right) \left((\eta+r\eta_d) + (\eta+s\eta_d)-(\eta+r\eta_d)(\eta+s\eta_d) \frac{\mathcal{R}}{\mathcal{R}_t^0}\right) \Big] \nonumber \\
&+& 4 (\eta+r\eta_d) (\eta+s\eta_d) R_{t1}^* R_{t1}  S_{t1}  \left((\eta+s\eta_d)   \frac{\mathcal{R}}{\mathcal{R}_t^0}-2 \right) 
 \left((\eta+s\eta_d) \frac{\mathcal{R}}{\mathcal{R}_t^0}-1\right)\nonumber \\
&+& \left. (\eta+r\eta_d) (\eta+s\eta_d) S_{t1}^* \left[ 2 R_{t1}
   \left((\eta+s\eta_d) \frac{\mathcal{R}}{\mathcal{R}_t^0}-1\right)
   + (\eta+s\eta_d) \frac{\mathcal{R}}{\mathcal{R}_t^0} S_{t1}+S_{t1}\right] \right\} \,,
   \label{L0}
\end{eqnarray}
where
\begin{equation}
R_{t1}=\frac{ib}{\mathcal{R}+ia}(S_{d1}+S_{n1})
\end{equation}
is defined  from the transitional Reynolds number $\mathcal{R}_t/\mathcal{R}_t^0=1-(k\zeta) e^{ikx} R_{t1}$ and deduced from Hanratty's relaxation relation (Eq. \ref{HanrattyRelaxation2}).

Introducing the above expansions into the Navier-Stokes equations (Eqs. \ref{NScont} - \ref{NSmomentum}), we derive the differential equations at linear and homogeneous-quadratic orders on the modal functions. The linear  {components yield}:
\begin{eqnarray}
U_1' &= & - i W_1 +\frac{S_{t1}-2  \Upsilon \mathcal{U}'^2 L_1}{{\mathcal{R}}^{-1}+2  \Upsilon^2 \mathcal{U}' }\,,\label{U1}
\\
W_1'&=&-iU_1\,,\\
\label{W1}
S_{t1}'&=&i\mathcal{U} U_1 +\mathcal{U}'W_1+i(S_{n1}+S_{d1})\,,\\
\label{St1}
S_{n1}'&=& -i\mathcal{U} W_1 + iS_{t1}\,,
\label{Sn1}
\end{eqnarray}
where the expression giving $U_1' $ (Eq. \ref{U1})  {results} from the shear stress modal function $S_{t1}$ (Eq. \ref{FunctionSt1}). This is the result derived in \citet{claudin2017dissolution}. For the new homogeneous-quadratic order, the differential equations  {of} the modal functions are:
\begin{eqnarray}
U_0'&=&\frac{1}{\left(2 \Upsilon^2 \mathcal{U}'+\mathcal{R}^{-1}\right)}\Bigg[ \Bigg.
S_{t0}-\frac{1}{2} L_1 \mathcal{U}'^2 L_1^*- \Upsilon^2
   \left( U_1 U_1^*+\frac{1}{2} (U_1'+i W_1) \left(U_1'^*-i W_1^*\right)\right)\Bigg. 
\nonumber
\\
&-&\Upsilon \left( \mathcal{U}' L_1^* (U_1'+i W_1)+ L_1 \mathcal{U}' \left(U_1'^*-i
   W_1^*\right)+2 L_0 \mathcal{U}'^2\right)   \Bigg]\,,  \label{U00}
   \\
S_{t0}'&=& \frac{1}{4}(W_1U_1'^*+W_1^*U_1')\,,
\label{St0}
\\
S_{n0}'&=&\frac{1}{2} i (U_1W_1^*-U_1^*W_1)\,,
\label{Sn0}
\end{eqnarray}
where, as in the linear order, the expression for $U_0'$ (Eq.~\ref{U00}) is deduced from the shear stress modal function $S_{t0}$ (Eq.~\ref{FunctionSt0}). Note that $S_{n0}$ does not enter the equations for $U'_{0}$ and $S'_{t0}$, and it thus decouples from the present problem focused on the computation of $U_0$.  {Furthermore,} the stress functions $S_{t0}$ and $S_{n0}$ can be obtained by integration over $\eta$, once the linear order ($U_1$ and $W_1$) is known.

The integration of the two above sets of differential equations (\ref{U1})-(\ref{Sn0}) requires boundary conditions at the substrate surface and at infinity. The upper boundary corresponds to the limit $\eta \to \infty$, in which the vertical flux of momentum vanishes asymptotically. This means that the corrections to the vertical velocity and to the shear stress must tend to zero at both orders: (i) $W_1(\infty)=0$ and (ii) $S_{t1}(\infty)=0$, $S_{t0}(\infty)=0$. In practice, we introduce a finite height $H$ (or $\eta_H \equiv kH$), at which we impose a null vertical velocity and a constant tangential stress $-\rho u_*^2$, so that $W_1(\eta_H) = 0$ and $S_{t1}(\eta_H) = 0$, $S_{t0}(\eta_H) = 0$. Then, we consider the limit $H \to +\infty$, i.e. when the results become independent of $H$. Both components of the velocity should vanish at the substrate surface, i.e. for $\eta=kZ$. The condition $u_x(x,kZ)=0$ then  {yields}: 
\begin{eqnarray}
U_1(0)  &=&  -\mathcal{U}'(0)\,,\\
\label{U1BC}
U_0(0)& = & -\frac{1}{4}\mathcal{U}''(0) - \frac{1}{4}U_1'(0) - \frac{1}{4}U_1'^*(0). 
\label{U0BC}
\end{eqnarray}
Similarly, for $u_z(x,kZ)=0$, we get $W_1(0) = 0$.

The differential equations (\ref{U1} - \ref{Sn1}) for the functions associated with the linear terms are first solved by integrating with the corresponding boundary conditions. The solutions are then used to solve equations (\ref{U00} - \ref{St0}) for the functions associated with the homogeneous terms. 

\section{Shifted representations and stream function}
\label{ShiftedRepresentationStreamFunction}

A `shifted' representation of the modal functions is used in some parts of the paper, noted with an additional tilde. In this study, all fields are expanded up to the homogeneous second order in $k\zeta$ in a non-shifted representation as in Eq. \ref{VelocityNonLinearExpansion} for the streamwise velocity. For some of the graphs, we sometimes prefer to present the velocity profiles in the shifted representation as
\begin{equation}
u_x= {u_*} \left[\mathcal U (\xi)+(k\zeta) e^{ikx} \tilde U_1(\xi )+(k\zeta)^2 \tilde U_0(\xi )\right]\,,
\label{uxns}
\end{equation}
where $\xi = \eta-k\zeta e^{ikx} $. Expanding the functions $\tilde{f} (\eta-k\zeta e^{ikx})$ with respect to $k\zeta$, one obtains 
\begin{equation}
\tilde U_0=U_0+\frac{1}{4}\mathcal U''+ \frac{1}{4}\left(U_1' +  U_1'^* \right)\,,
\label{u0ns}
\end{equation}
where $\tilde U_1 =U_1+\mathcal U'$ has been given in \citet{fourriere2010bedforms}. The shifted representations for the other fields work in a similar fashion, except that there could be no base profile as $\mathcal U$ in the expressions.

To compute the streamlines, we introduce the stream function $\Psi(x,z)$. In the shifted representation, one solution is $\Psi=\int _0^{z}  u_x {\rm d} \hat{z}$, and this integral is computed between the surface $\hat{z} = 0$ and some vertical distance $\hat{z} = z$. Following Eq. \ref{uxns}, one then has
\begin{equation}
\Psi=\frac{u_*}{k}\int_0^{\xi} \left[\mathcal U (\hat{\xi})+(k\zeta) e^{ikx} \tilde U_1(\hat{\xi} )+(k\zeta)^2 \tilde U_0(\hat{\xi} )\right]{\rm d} \hat{\xi}\,.
\label{Psi1}
\end{equation}
Introducing $\tilde U_1 =U_1+\mathcal U'$ and Eq. \ref{u0ns}, it gives
\begin{equation}
\Psi=\frac{u_*}{k}\int_0^{\eta} \left[\mathcal U +(k\zeta)^2 \left( U_0+\frac{1}{4}\mathcal U''+ \frac{1}{4}\left(U_1' +  U_1'^* \right)\right) +(k\zeta) e^{ikx} \left( i W'_1+ \mathcal U' \right)\right]{\rm d} \hat{\eta}\,,
\label{Psi2}
\end{equation}
%
where we have used the relation at the linear order $W_1'=-iU_1$ (Eq. \ref{W1}). Consequently, one obtains a dimensionless stream function $k\Psi/u_*$
\begin{eqnarray}
\frac{k \Psi}{u_*}&=&\int_{0}^{\eta} \left(\mathcal U + (k\zeta)^2 U_0 \right){\rm d} \hat{\eta} + \frac{1}{4} (k\zeta)^2 \left[\mathcal U'+ \left(U_1 +  U_1^* \right)+ \mathcal U'(0) \right] \nonumber\\
&+&(k\zeta) e^{ikx}\left( i W_1+ \mathcal U \right),
\label{Psi3}
\end{eqnarray}
where we have used the boundary conditions $\mathcal U(0) =0$, $W_1(0)=0$ and $U_1(0) = -\mathcal U'(0)$.
%
%
\textcolor{black}{ For the stream function in the self - consistent expansion analysis (section \ref{LoopedExpansion}), it should be noted that in this equation the homogeneous velocity of the base state $\mathcal U$ in the integration should be estimated with $\mathcal R_d = 0$ as $\mathcal U^0$, and that the rest quantities should be estimated with $\mathcal R_d = d_e u_*/\nu$.
%
}
Here, for the plots (figure~\ref{kz0_Rdfit}c), only the real part of (\ref{Psi3}) is considered.

\section{Asymptotic regimes of the roughness coefficient $E$}
\label{AsymptoticRegimes}

 {In this appendix, we provide more details in the scaling arguments for the viscous and turbulent regimes of the roughness coefficient $E$.}
\subsection{Viscous regime}
In the viscous response limit, the flow is governed by the Stokes equations. At the linear order in $k\zeta$, the corresponding equations on the modal functions are
\begin{eqnarray}
U_1' &= & - i W_1 +\frac{S_{t1}}{{\mathcal{R}}^{-1} }\,,
\label{SU1vis}\\
W_1'&=&-iU_1\,,\\
\label{SW1vis}
S_{t1}'&=& i(S_{n1}+S_{d1})\,,
\label{SSt1vis}\\
S_{n1}'&=&  iS_{t1}\, .
\label{SSn1vis}
\end{eqnarray}
At the homogeneous quadratic order, we have
\begin{eqnarray}
U_0'&=&\frac{S_{t0}}{\mathcal{R}^{-1}}\,,
\label{U0vis} \\
S_{t0}'&=& 0\,,
\label{St0vis} \\
S_{n0}'&=&0\,.
\label{Sn0vis} 
\end{eqnarray}
Using the boundary condition $S_{t0}(\eta_H)=0$, we can deduce from (\ref{St0vis}) that $S_{t0}  = 0$. From (\ref{U0vis}), we therefore obtain $U'_0 = 0$, i.e. $U_0$ is a constant. Considering the  {no-slip} boundary condition at the surface (Eq. \ref{U0BC}), we have 
\begin{equation}
U_0=U_0(0)  =  -\frac{1}{4}\mathcal{U}''(0) - \frac{1}{4}U_1'(0) - \frac{1}{4}U_1'^*(0)\,.
\label{U0vis2}
\end{equation}
In this viscous limit $\mathcal{U}' = \mathcal{R}$, independent of $\eta$, and therefore $\mathcal{U}''(0) = 0$\,.  From (\ref{SU1vis}), we deduce 
\begin{equation}
U_1'(0) = {S_{t1}(0)}{{\mathcal{R}}}\,,
\label{U1primevis}
\end{equation}
where we have used the condition $W_1 (0) = 0$. Considering $S_{t1}(0) =\mathcal A + i \mathcal B$, we then have 
\begin{equation}
U_0=  - \frac{1}{4}(S_{t1}(0) + S_{t1}^*(0)) = -  \frac{1}{2} \mathcal{R} \,\mathcal A \,,
\label{U0vis3}
\end{equation}
and thus
\begin{equation}
E = - U_0 = \frac{1}{2} \mathcal{R} \, \mathcal A \,.
\label{Evis}
\end{equation}

 {
\subsection{Turbulent regime}
In this limit, one can infer an expression for $E(kz_0)$ estimating $U_0$ at a typical (dimensionless) height $\eta \sim 1$, above which its profile becomes constant (figure\ref{U0_eta_Rd}a). For that purpose, we combine estimations of the shear stress from the momentum balance (\ref{NSmomentum}) and from the turbulent closure  (\ref{StrainStress}). 

Let us start with (\ref{StrainStress}), we can express in this limit as $\tau \sim \rho \ell^2 \left( \nabla u \right)^2$. For the homogeneous correction in $(k\zeta)^2$, it writes
\begin{equation}
S_{t0} = \Upsilon^2 \mathcal{U}' U_0' + L_0^2 \mathcal{U}'^2 + \Upsilon^2 U_1^2 + \Upsilon^2 U_1'^2 + \Upsilon L_1 \mathcal{U}' U_1' + L_1^2 \mathcal{U}'^2
\label{StrainStressTurbulentOrder0}
\end{equation}
where the first term on the RHS comes from the velocity gradient taken at the quadratic order $U_0$, with all other factors taken in the base state ($\Upsilon$ for the dimensionless mixing length and $\mathcal{U}$ for the dimensionless velocity). The second term is the modulated mixing length at this order 0, and the base velocity. The other contributions come from combinations of terms at linear order. A first possibility is to take a base mixing length with velocity horizontal and vertical derivatives. A second possibility is with a modulated mixing length and (vertical) velocity derivative. A last possibility is to take a modulated mixing length with a base velocity. One can recognise all these terms in Eq.~\ref{FunctionSt0}, properly expressed with complex notations and correct prefactors. We can estimate their scaling behaviour having in mind that, in this rough turbulent limit, we have: $\Upsilon \sim \kappa \eta$, $\mathcal{U}' \sim 1/\Upsilon$, $L_1 \sim \kappa$ and $L_0 \to 0$. Also, we estimate quantities and vertical derivatives at the scale $\eta \sim 1$, so that $\mathcal{U}' \sim \mathcal{U}/\eta \sim \mathcal{U}$ (and similarly for $U_1$ and $U_0$). It essentially leads to $S_{t0} \sim U_0 + U_1^2 + U_1 + \mbox{Cst}$.
}

\backsection[Acknowledgements]{We thank F. Charru, O. Dur{\'a}n~Vinent and M. Louge for long term collaborations on this subject of hydrodynamic response to surface topography. We acknowledge the contribution of A. Fourri\`ere in setting a first version of these non-linear hydrodynamic calculations. P.C. thanks W. Anderson, S. Carpy, S. Courrech du Pont, L. Couston, L. Duchemin, B. Favier and C. Narteau for discussions on Hanratty's anomaly.}

\backsection[Funding]{Pan Jia was supported by Shenzhen Science and Technology Programme (Grant No. 629 RCBS20200714114940144), and National Natural Science Foundation of China (NSFC12102114). }

\backsection[Declaration of interests]{The authors report no conflict of interest.}

\backsection[Data availability statement]{The data that support the findings of this study are available from the authors upon request.}

\backsection[Author ORCID]{Pan Jia, https://orcid.org/0000-0002-3988-4363; Bruno Andreotti, https://orcid.org/0000-0001-8328-6232; Philippe Claudin, https://orcid.org/0000-0001-8975-4500.}


\bibliographystyle{jfm}
\bibliography{jfm}

%
%
%
%
%
%
%
%
%
%

\end{document}